\documentclass[conference]{IEEEtran}

\ifCLASSOPTIONcompsoc
  \usepackage[nocompress]{cite}
\else
  \usepackage{cite}
\fi
\usepackage[utf8]{inputenc}
\usepackage[usenames, dvipsnames]{color}
\usepackage{boldline}

\ifCLASSINFOpdf
\usepackage[pdftex]{graphicx}
\graphicspath{{Figures/}}
\DeclareGraphicsExtensions{.pdf,.jpeg,.png}
\usepackage{subfig}
\usepackage{tabularx}
\usepackage{url}
\usepackage{amsmath}
\begin{document}

\title{Intelligent Middle-Level Game Control}

\author{\IEEEauthorblockN{Amin Babadi}
\IEEEauthorblockA{\textit{Department of Computer Science}\\
\textit{Aalto University}\\
Helsinki, Finland\\
amin.babadi@aalto.fi}
\and
\IEEEauthorblockN{Kourosh Naderi}
\IEEEauthorblockA{\textit{Department of Computer Science}\\
\textit{Aalto University}\\
Helsinki, Finland\\
kourosh.naderi@aalto.fi}
\and
\IEEEauthorblockN{Perttu Hämäläinen}
\IEEEauthorblockA{\textit{Department of Computer Science}\\
\textit{Aalto University}\\
Helsinki, Finland\\
perttu.hamalainen@aalto.fi}
}
\maketitle



\begin{abstract}
We propose the concept of \textit{intelligent middle-level game control}, which lies on a continuum of control abstraction levels between the following two dual opposites: 1) high-level control that translates player's simple commands into complex actions (such as pressing Space key for jumping), and 2) low-level control which simulates real-life complexities by directly manipulating, e.g., joint rotations of the character as it is done in the runner game QWOP. We posit that various novel control abstractions can be explored using recent advances in movement intelligence of game characters. We demonstrate this through design and evaluation of a novel $2$-player martial arts game prototype. In this game, each player guides a simulated humanoid character by clicking and dragging body parts. This defines the cost function for an online continuous control algorithm that executes the requested movement. Our control algorithm uses Covariance Matrix Adaptation Evolution Strategy (CMA-ES) in a rolling horizon manner with custom population seeding techniques. Our playtesting data indicates that intelligent middle-level control results in producing novel and innovative gameplay without frustrating interface complexities.
\end{abstract}


%
\IEEEpeerreviewmaketitle

\begin{IEEEkeywords}
game control, physically-based simulation, multi-agent systems, online optimization, continuous control
\end{IEEEkeywords}


\section{Introduction}

\begin{figure*}[htbp]
    \centering
    \subfloat[Task = \textit{Null}]{{\includegraphics[width=0.33\linewidth]{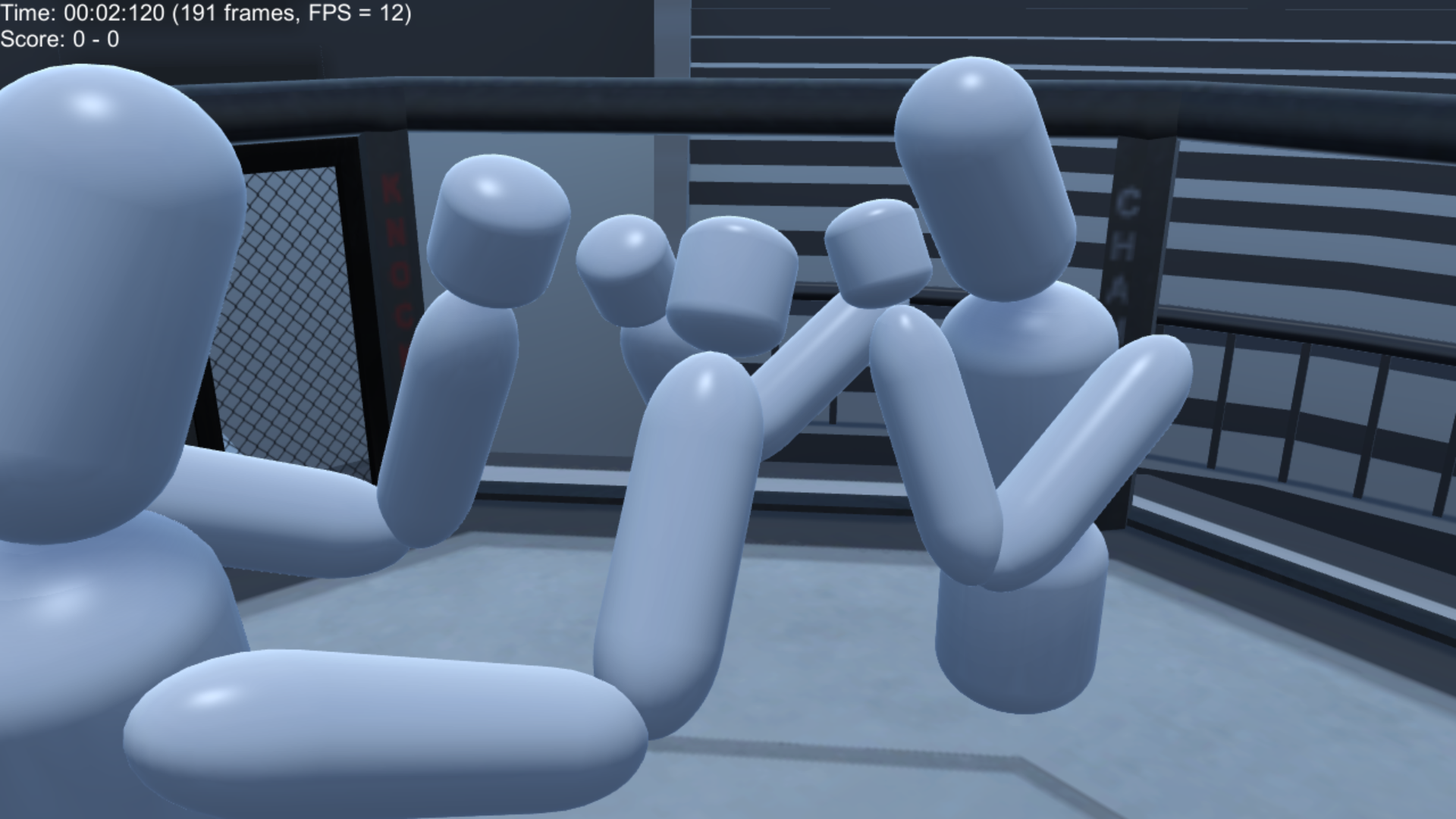} }\label{Fig-TwoPlayerGameScreenshot-A}}
    \subfloat[Task = \textit{Punch in head using right hand}]{{\includegraphics[width=0.33\linewidth]{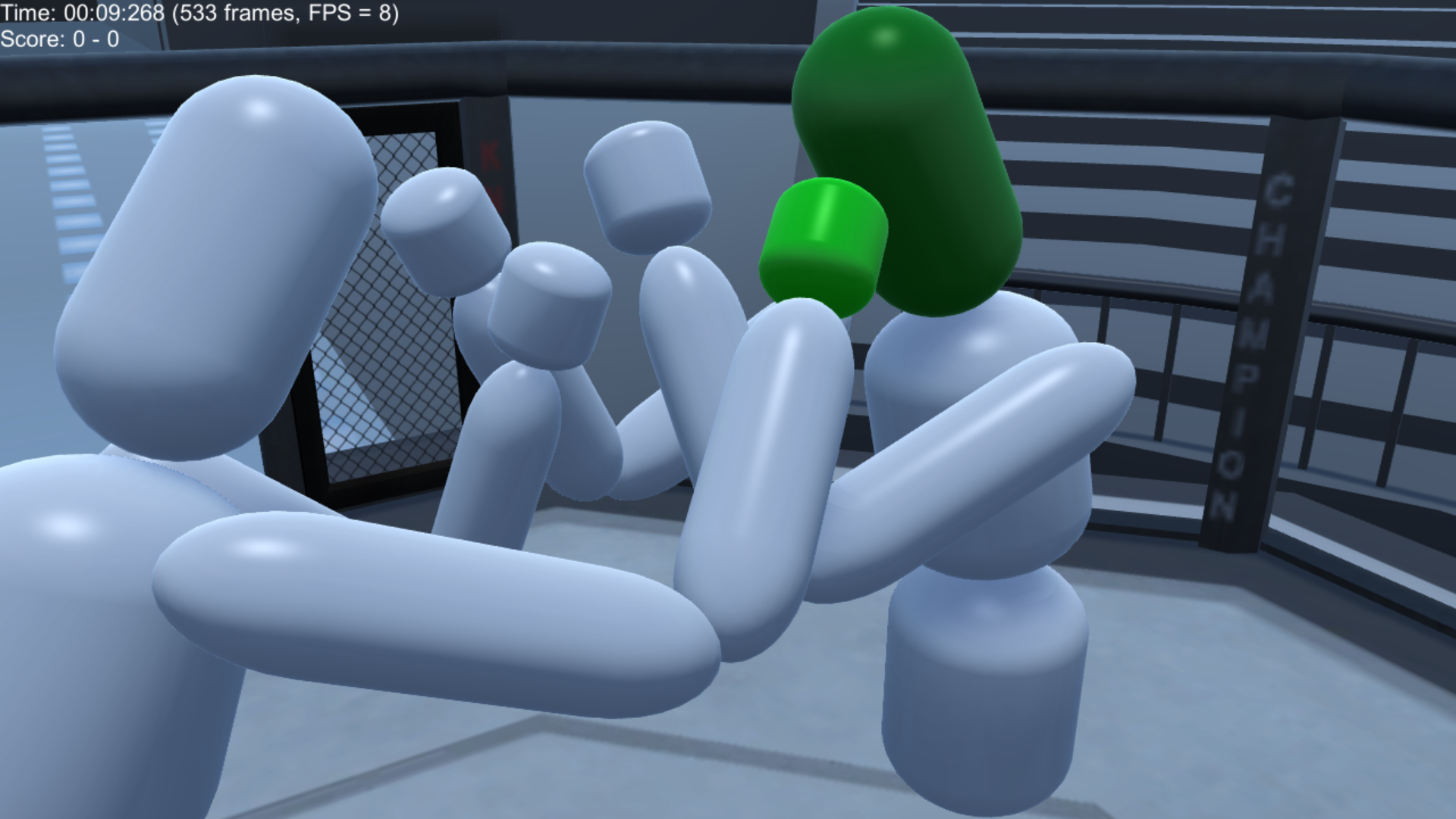} }\label{Fig-TwoPlayerGameScreenshot-B}}
    \subfloat[Task = \textit{Move right hand to specified position}]{{\includegraphics[width=0.33\linewidth]{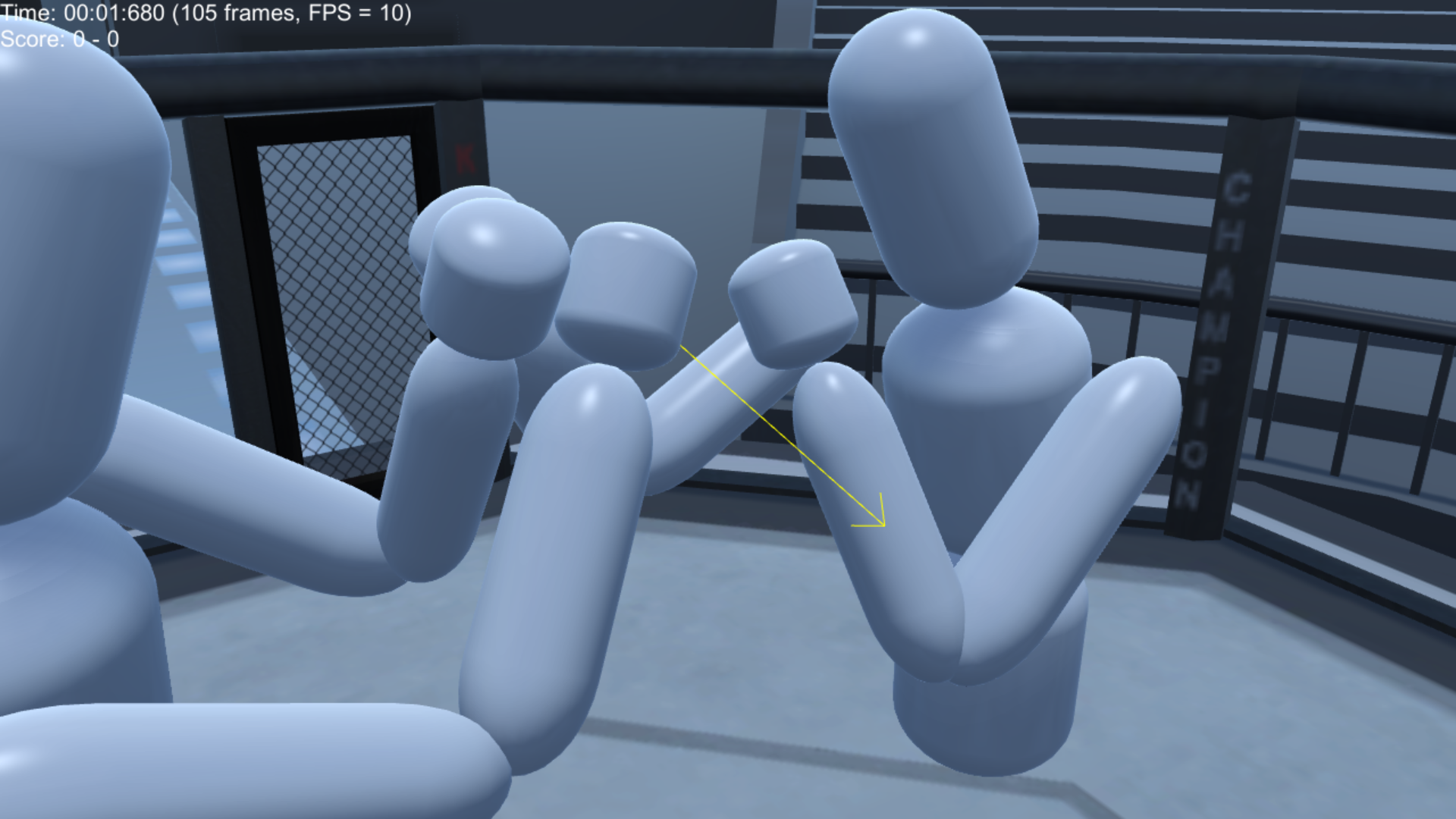} }\label{Fig-TwoPlayerGameScreenshot-C}}
    \caption{Screenshots of our $2$-player martial arts game prototype with $3$ different tasks. The left character is controlled by the local player.}
    \label{Fig-TwoPlayerGameScreenshot}
\end{figure*}

Game control in present video games can be divided into \textit{high-level} and \textit{low-level} control \cite{Hamalainen2017Predictive}. In high-level control, the player is able to trigger actions such as punch or kick with a simple keypress. In contrast, there are low-level control approaches where the player directly manipulates the game system simulation. For example, using this approach in a martial arts game could mean that the player has to determine the torques applied to each joint of the character's body to produce a punch. The fighting game Toribash\cite{toribash} and the runner game QWOP\cite{qwop} are two of the games that have successfully used this kind of low-level control.

Low-level control allows maximal expressiveness, diversity/complexity, and control in game animations; it can also remove the cost of animation production in game projects. However, it usually makes the character control extremely difficult since it requires the player to manipulate several degrees of freedom with high precision, often under time pressure. On the other hand, games with high-level control usually come with a set of pre-defined smooth and natural animations. The downside is that  the animations are costly to produce, and they do not allow for interesting novel movements to emerge \cite{Hamalainen2017Predictive}.

In this paper, we propose \textit{intelligent middle-level game control} by combining usability and flexibility of high-level and low-level controls, respectively. Using this approach, the player's commands are more abstract and simple than low-level control, but more detailed and expressive than high-level control. To make this possible, we utilize recent movement artificial intelligence (AI) techniques for physically-simulated characters. To clarify this definition, consider the martial arts game example again. Suppose the player can produce the command \textit{"use your left hand to push the opponent's right hand away"} by left-clicking on the opponent's right hand. Then, the game automatically computes and then applies the required torques for producing the requested animation, adapting to the current physical state of the characters. In other words, player commands cause the animations to be generated on the fly and no pre-recorded animations are used. The main advantages of intelligent middle-level control are as follows:
\begin{enumerate}
\item The synthesized movements are novel and emergent similar to low-level control, but the player can focus on more strategic planning instead of micro-managing the simulation.
\item The complexity of a simulated human body's dynamics can create interesting challenges \cite{Hamalainen2017Predictive}. Both novelty and complexity are desirable from the point of view of inducing a feeling of curious interest in the player \cite{silvia2008interest}. 
\item Since movement is not limited to pre-defined animations, more expressive and precise game controls can be designed and implemented.  
\end{enumerate}

Middle-level control has been explored before in a few games such as Octodad \cite{octodad} and the original PC version of Rag Doll Kung Fu \cite{ragdollkungfu}, albeit with less control intelligence. The games typically use some form of inverse kinematics which limits the behaviors that can be created. We demonstrate that intelligent middle-level control with online trajectory optimization allows realistic handling of physical constraints such as joint limits and non-penetration of colliding bodies. 


We believe that intelligent middle-level control has the potential for introducing various novel gameplay. To support this claim, we developed a novel $2$-player martial arts game prototype in which the players are able to control their physically-based humanoid characters through giving middle-level commands. We also developed an online continuous control trajectory optimization algorithm that computes the simulation control parameters needed to produce the requested movements. Screenshots of the game are shown in Fig. \ref{Fig-TwoPlayerGameScreenshot} and a gameplay video is available online\footnote{\url{https://youtu.be/rnsSWY7HZJA}}. We evaluated this prototype by running a user study with $12$ participants ($6$  human-vs-human pairs). The players reported that the interface enabled them to use various martial arts strategies, and that low-level controller was able to produce their commands with high precision.

The rest of the paper is organized as follows: A brief overview of literature is given in Section \ref{Sec-RelatedWorks}. Section \ref{Sec-Method} explains the details of intelligent middle-level control and our martial arts game prototype. Section \ref{Sec-Evaluations} describes the details and results of the user study that was run for evaluating intelligent middle-level control in our game. Finally, Section \ref{Sec-Conclusions} gives conclusions and Section \ref{Sec-FutureWork} analyzes the limitations and future lines of research in this work.


\section{Related Works} \label{Sec-RelatedWorks}
In this section, we first give an overview of recent methods for synthesis and control of physically-based character animation, followed by game control interface research relevant to this work.


\subsection{Character Movement Synthesis}\label{SubSec-CharacterMovementSynthesis}
Traditional animation technology has limited movement expressiveness and emergence, except for simple low-level simulation control (e.g., Toribash\cite{toribash} and QWOP\cite{qwop}). However, this is changing due to deep reinforcement learning and novel real-time movement optimization methods, which can endow game characters with expressive movement intelligence not limited to pre-defined animations.  We refer the reader to \cite{geijtenbeek2011interactive} for a thorough introduction to character animation and physically-based simulation along with a survey on common techniques.

\subsubsection*{Simulation-Based Methods}
In physical environments, behavior of objects and their interactions is usually difficult to model and predict. One of the most common approaches for character control in these environments is to use simulation-based methods. The basic idea behind these methods is simple: generate a number of action sequences, evaluate them using forward simulation and computing some cost function, and finally, choose the action sequence that minimizes the cost function.

If the simulation has differentiable dynamics, one can use dynamic programming version of gradient-based optimization \cite{tassa2012synthesis} to control a variety of systems ranging from an inverted pendulum to a full humanoid. With black-box simulation, similar results were obtained by Sequential Monte Carlo sampling of control trajectories encoded as cubic splines \cite{hamalainen2014online}. Instead of a spline parameterization, Control Particle Belief Propagation (C-PBP) uses a Markov Random Field factorization for both sampling and smoothing trajectories \cite{hamalainen2015online}. Rajamäki and Hämäläinen \cite{rajamaki2017augmenting} have recently shown that adding supervised learning on top of Monte Carlo tree search (MCTS) methods \cite{browne2012survey} can yield both robust control and low movement noise.

Several simulation-based methods have been developed using evolutionary computation. A recent study has used graph search along with Covariance Matrix Adaptation Evolution Strategy (CMA-ES) \cite{hansen2006cma} to develop an offline controller for solving humanoid wall climbing problems \cite{naderi2017discovering}. It has been shown that CMA-ES can be used in a rolling horizon manner in single-agent control problems with continuous states and actions \cite{samothrakis2014rolling} and $2$-player games with discrete actions \cite{liu2016rolling}. Another study has shown that performance of rolling horizon evolutionary algorithms can be improved significantly using simple population seeding techniques \cite{gaina2017population}. We are investigating this approach in the context of a more complex $2$-player simulation with continuous actions and complex contact dynamics. 

\subsubsection*{Reinforcement Learning}
Reinforcement learning (RL) is a field of machine learning that studies how an agent should take actions in an environment in order to maximize rewards. In the past few years, RL has received a lot more attention due to remarkable results of Deep Reinforcement Learning (DRL) in Atari games \cite{mnih2015human} and the game of Go \cite{silver2016mastering, silver2017mastering}. These advances have inspired several  breakthroughs in RL for continuous control. Mixture of actor-critic experts (MACE) accelerates learning by developing separate pairs of actors and critics such that each pair learns some part of the movement \cite{peng2016terrain}. Schulman \textit{et al.} \cite{schulman2015trust} introduced a method called Trust Region Policy Optimization (TRPO) that uses a surrogate objective function and is able to learn several complex tasks such as swimming and walking. Another method called Proximal Policy Optimization (PPO) has managed to outperform TRPO by clipping the surrogate objective function \cite{schulman2017proximal}. The most important limitation of DRL methods is that they need a lot of simulation data and training time to learn. This can be a problem especially when iterating the reward function design.

\subsubsection*{Data-Driven Methods}
Studies have shown that data-driven methods can be effective for generating robust and smooth movements. One of the studies has shown that kinematic controllers can be constructed by learning a low-dimensional space from motion capture data and interpolating in that space \cite{levine2012continuous}. Motion matching is a similar kinematic method that uses a dataset of pre-recorded animations and in each frame finds the closest pose to the character's current pose such that desired future movement is produced \cite{clavet2016motion}. Holden \textit{et al.} \cite{holden2016deep} use convolutional autoencoders on a large motion capture data set to re-produce and interpolate recorded motions. Another study breaks control problem into short time fragments ($0.1$s in length), and learns a linear feedback control strategy for each fragment \cite{liu2016guided}. A more recent method, called DeepLoco, uses a combination of high-level and low-level controllers and is able to produce stable gaits given some reference motions \cite{peng2017deeploco}. Phase-functioned neural network is a recent neural network architecture that uses cyclic functions for computing the weights and is trained using a large dataset of pre-recorded animations \cite{holden2017phase}. Although data-driven methods are able to produce high-quality movements, it can be difficult to obtain the training data, and the resulting movement is limited by the data.

\subsection{Game Control Interface}\label{SubSec-GameControlInterface}
There is a lot of research on alternative input devices and interfaces, e.g., for controlling games with body movements \cite{Krome:2017:AEG, Merritt:2017:GDP, Kajastila2016theaugmented}. On the other hand, some studies have proposed using traditional input devices but novel input-avatar mappings \cite{sheinin2015quantifying, Hamalainen2017Predictive}.

A recent study has used predictive simulation for developing $6$ novel game prototypes using low-level control of physically-based simulated characters \cite{Hamalainen2017Predictive}. Our work is close to this work as we solve the same problem of enabling expressive control of fully physically-simulated characters without excessive control difficulty. They propose predictive visualizations as the solution, whereas we explore the possibility of offloading some complexity onto the movement optimizer.

It can be argued that middle-level control is not an entirely new concept. For example, games like Octodad \cite{octodad}, and the original PC version of Rag Doll Kung Fu \cite{ragdollkungfu} have used inverse kinematics style control. Compared to Toribash \cite{toribash} and QWOP \cite{qwop}, they provide the player with a slightly higher-level interface of directly controlling hand and feet locations instead of joint rotations. Our contribution is in demonstrating how modern physically-based optimization methods for online continuous control provide novel tools for exploring game control interfaces and abstractions.


\section{Method} \label{Sec-Method}
In this section we describe our $2$-player martial arts game prototype demonstrating the intelligent middle-level control concept. At Section \ref{SubSec-CharacterModel}, the details of reference model are explained. Then, at Section \ref{SubSec-MidLevelControl} we explain the design of intelligent middle-level control in our martial arts game prototype. At Section \ref{SubSec-LowLevelController}, the structure of low-level controller is introduced. Finally, the network architecture of our martial arts game is explained in Section \ref{SubSec-NetworkArchitecture}. The values of all parameters introduced in this section are reported at Appendix.

\subsection{Character Model}\label{SubSec-CharacterModel}
\begin{figure}[htbp]
\centerline{\includegraphics[width=.6\linewidth]{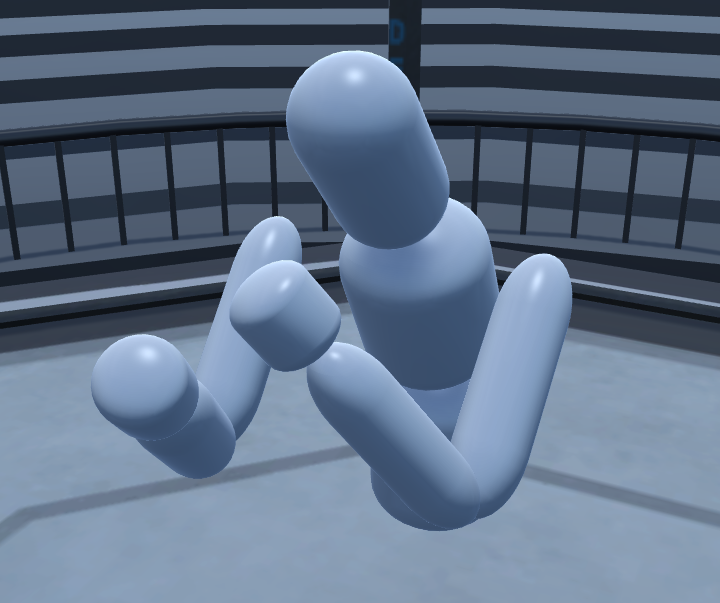}}
\caption{Upper-body character model in its reference martial arts pose.}
\label{Fig-CharacterModel}
\end{figure}

Characters in this game are upper-body humanoid characters with $16$ actuated degrees of freedom (DOF) as shown in Fig. \ref{Fig-CharacterModel}. Each character has $9$ bones that are connected using $3$-DOF ball-and-socket or $1$-DOF hinge joints. We use Open Dynamics Engine (ODE) \cite{ode} for physics simulations.

\subsection{Novel Middle-Level Control for Martial Arts Games}\label{SubSec-MidLevelControl}
\begin{figure}[htbp]
\centerline{\includegraphics[width=.7\linewidth]{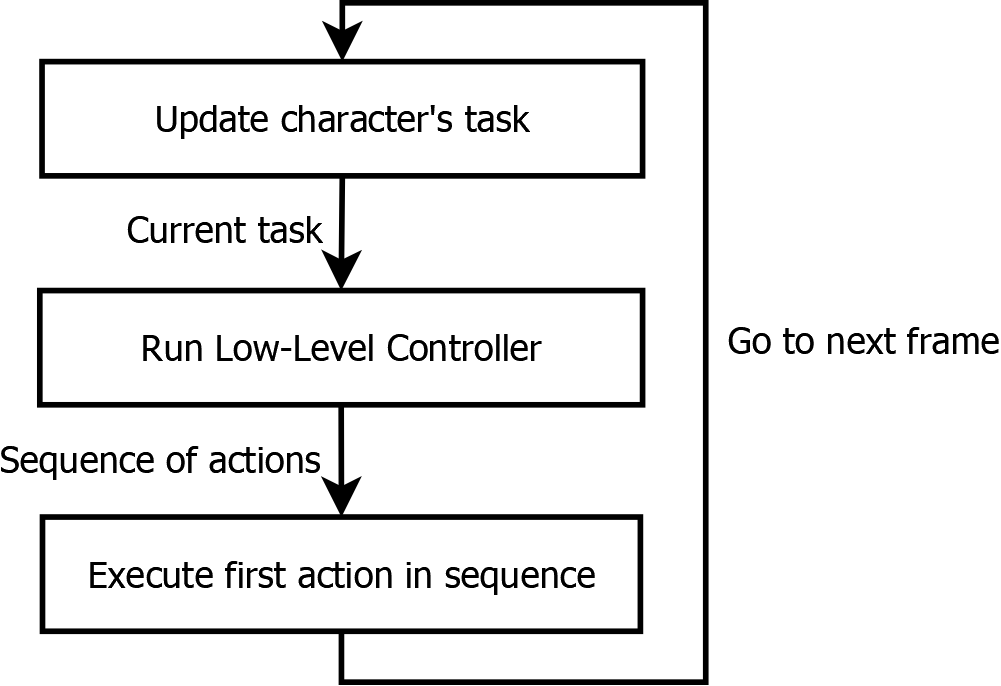}}
\caption{Core loop in intelligent middle-level control.}
\label{Fig-MidLevelLoop}
\end{figure}

The overall schema of our control system is shown in Fig. \ref{Fig-MidLevelLoop}. In each frame, the current task of the character is determined and then a low-level controller plans a sequence of actions in order to complete the task. Then, the first action in the sequence is executed and the simulation goes to the next frame. This form of online optimization, i.e., 1) finding a multi-step solution, 2) applying the initial step of the solution, and 3) moving/rolling the horizon forward, is called the rolling horizon control (also known as receding horizon control) \cite{chang2007simulation}. Note that the character is not necessarily given a new task in each frame in which case the task from previous frame is considered as the current task.

\subsubsection*{Tasks} Characters can have $2$ main tasks: 1) \textit{Move} hand to a specific position, and 2) \textit{Punch} opponent either in the head or in the chest. The \textit{Move} task is enabled using mouse drag in which case the desired position is determined based on hand's current position and mouse drag direction. A single click on opponent's head or chest enables the \textit{Punch} task for those parts. Both tasks can be executed using either hands depending on the clicked mouse button (i.e., the left click for the left hand and the right click for the right hand). There is also a dummy \textit{Null} task defined for specifying a character with no tasks. Character's task is set to \textit{Null} if 1) current task is completed, or 2) time spent on task has passed some threshold $\tau_{\textit{Max Task}}$.

\subsubsection*{Heads-Up Display (HUD) for tasks} The current task of character is indicated using a simple color coding as follows: 1) when the character is given a punching command, the operating hand and punching target are highlighted using green color, 2) when a successful punch is executed, the punching target will become red, and 3) when a moving command is set for one of the hands, a yellow line connects the current position of the hand to its desired position.

\subsubsection*{Additional input keys} Each player can use W/S keys to move his/her character on a horizontal line. We added this ability because in martial arts it is very important for the characters to adjust distance to their opponent. Finally, each player can rotate the camera around his/her character using A/D keys.

\subsection{Low-Level Controller}\label{SubSec-LowLevelController}
In each frame, we need a low-level controller for achieving the character's current task. To this end, we developed an online simulation-based algorithm for physically-based continuous control. This algorithm uses CMA-ES in a rolling horizon manner along with custom population seeding techniques and outputs the best found action sequence up to a fixed horizon $\tau_{\textit{Horizon}}$. CMA-ES is a common evolutionary algorithm that assumes a multi-variate Gaussian distribution as the underlying data distribution. In each iteration of CMA-ES, a new population is generated using some assumed distribution. Then, after evaluating the fitness of each individual, CMA-ES updates mean and covariance matrix of Gaussian distribution by selecting a subset of population with highest fitness values \cite{hansen2006cma}. Population seeding means that in each iteration of CMA-ES, some proportion of the population is generated using external distributions. To the best of our knowledge, this is the first time that CMA-ES is used in a rolling horizon manner for online control of 3D physically-based simulated characters.

We first tried using the combined tree search and supervised learning approach of Rajamäki and Hämäläinen \cite{rajamaki2017augmenting}, but found that it had difficulties generating extreme dynamic movements such as punches. We then tried our present approach, as the combination of CMA-ES and a spline parameterization was successfully utilized in the dynamic climbing movement synthesis of Naderi \textit{et al.} \cite{naderi2017discovering}. They however use CMA-ES for offline optimization instead of in online rolling horizon manner. 

\subsubsection*{Overall Search Method} In each frame, we run CMA-ES update $n_{\textit{CMA-ES Updates}}$ times. In each update, a population of size $n_{\textit{Population Size}}$ is generated using mean and covariance of CMA-ES and $2$ seeding techniques. Then the fitness value of each population member is computed and the CMA-ES updates its mean and covariance. After repeating this process $n_{\textit{CMA-ES Updates}}$ times, first action in the best found action sequence is returned as the character's next action.

\subsubsection*{Spline Parameterization} We parameterize each action sequence using a cubic spline of $n_{\textit{Spline Points}}=3$ control points. This reduces the problem dimension significantly since we do not need to optimize each individual action in the sequence.  Plus, interpolation between control points enforces coordination between body joints which results in smooth movements. We also optimize the time variable of each control point; so each spline is defined using $3\times\left(16+1\right) = 51$ parameters.

\subsubsection*{Population Seeding Techniques} In each CMA-ES update, a fixed number of splines are generated using a multi-variate Gaussian distribution. The standard deviation of this distribution is $\sigma_{\textit{Pose}}$ and the mean is determined based on the following $2$ seeding techniques: 1) $n_{\textit{Last Best}}$ splines are generated by using the best spline found in the last frame as the mean. Note that at first CMA-ES update of each frame, we need to shift the last best spline by one frame so it becomes valid in the current frame. 2) $n_{\textit{Default Pose}}$ splines are generated by using the default martial arts pose shown in Fig. \ref{Fig-CharacterModel} as the mean.

\subsubsection*{Fitness Computation}
All action splines are evaluated in parallel using forward simulation up until some horizon $\tau_{\textit{Horizon}}$ by assuming time step of $\Delta t=1/30$ seconds and computing the reward (negative cost) in each time step. At the end of forward simulation, the fitness value of each action spline is equal to the average fitness value of all visited states during its simulation. The fitness value of a state $s$ is computed by summing over the negation of $3$ cost components as follows (the goal is to maximize the fitness):

\begin{equation}
\begin{split} 
\textit{Fitness}\left(s\right) = - \left(\textit{Cost}_{\textit{Pose}}\left(s\right) + \textit{Cost}_{\textit{Move}}\left(s\right) + \textit{Cost}_{\textit{Punch}}\left(s\right)\right)
\end{split}
\end{equation}
where values of cost components $\textit{Cost}_{X}\left(s\right)$ are computed as follows:

\begin{enumerate}
\item $\textit{Cost}_{\textit{Pose}}\left(s\right)$: Cost of pose deviation is computed by finding the angle between current rotation of each bone ($q_{\textit{current}}^b$) and its desired rotation ($q_{\textit{desired}}^b$) as shown in the reference martial arts pose in Fig. \ref{Fig-CharacterModel}. The cost is computed as:

\begin{equation}
\begin{split} 
\textit{Cost}_{\textit{Pose}}\left(s\right)= \sum_{b}\left(\frac{\angle\left(q_{\textit{current}}^b, q_{\textit{desired}}^b\right)}{\sigma_{\textit{Pose}}}\right)^2
\end{split}
\end{equation}

where $\sigma_{\textit{Pose}}$ is used for indicating how much difference in rotation can be tolerated for each bone.

\item $\textit{Cost}_{Move}\left(s\right)$: Cost of moving hand $h$ is simply defined using the distance between current position $p_{\textit{current}}^h$ and desired position $p_{\textit{desired}}^h$ of the hand as follows:

\begin{equation}
\begin{split}
\textit{Cost}_{Move}\left(s\right)=\left(
\frac{\| p_{\textit{current}}^h - p_{\textit{desired}}^h \|_2}{\sigma_{\textit{Move}}}
\right)^2
\end{split}
\end{equation}

where $\sigma_{\textit{Move}}$ is used for tuning the amount of tolerance for difference between current and desired positions.

\item $\textit{Cost}_{\textit{Punch}}\left(s\right)$: Punching is the most complicated cost component in our work. In a good punch, the hand touches the target with highest possible speed. Then the hand should get back to its relaxed position quickly so the character's guard is not down for a long time. In order to favor these movements, punching cost is computed as follows:

\begin{equation}
\begin{split}
&\textit{Cost}_{\textit{Punch}}\left(s\right)= \\
&\textbf{1}_{\textit{punch not happened?}} \cdot \left(
\frac{\| v_{\textit{current}}^h - v_{\textit{desired}}^h \|_2}{\sigma_{\textit{Hand Velocity}}}
\right)^2 - \\
&\textbf{1}_{\textit{punch happened now?}} \cdot \textit{PunchPower}\left(v_{\textit{current}}^h\right) + \\
&\textbf{1}_{\textit{punch happened before?}} \cdot \left(
\frac{\| p_{\textit{current}}^h - p_{\textit{desired}}^h \|_2}{\sigma_{\textit{Hand Relax Position}}}
\right)^2
\end{split}
\end{equation}
where $\textbf{1}_A$ is the indicator function and is equal to $1$ if the condition $A$ is true, and $0$ otherwise. Current and desired velocity of hand are denoted by $v_{\textit{current}}^h$ and $v_{\textit{desired}}^h$, respectively. Similarly, $p_{\textit{current}}^h$ and $p_{\textit{desired}}^h$ denote current and desired position of the hand, respectively. Similar to previous cost components, $\sigma_{\textit{Hand Velocity}}$ and $\sigma_{\textit{Hand Relax Position}}$ determine the amount of tolerance for difference between current and desired values of the hand velocity and position, respectively. The function call $\textit{PunchPower}\left(v_{\textit{current}}^h\right)$ maps current velocity of the hand $h$ to a number in the range $[1000, 3000]$. 
\end{enumerate}

\subsection{Network Architecture}\label{SubSec-NetworkArchitecture}
\begin{figure}[htbp]
\centerline{\includegraphics[width=1.0\linewidth]{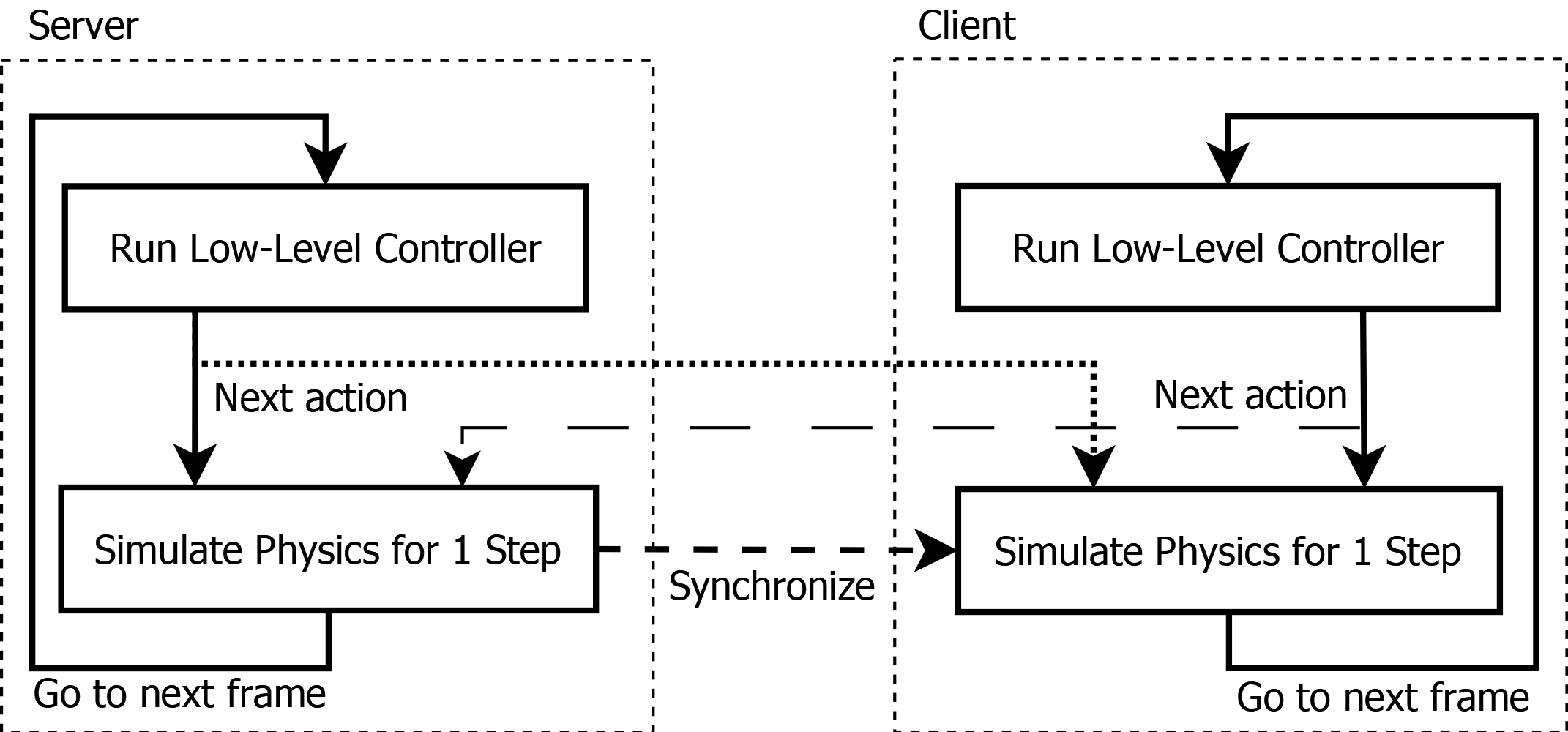}}
\caption{Network architecture and interactions between server and client.}
\label{Fig-NetworkArchitecture}
\end{figure}

One of the challenges for developing this prototype was how to design the network architecture. A critical limitation of previous multi-agent studies is that they mostly use competitive self-play RL, which is very slow and unreliable to train. We decided to run optimization in an interleaved manner to double our computing power. Our architecture is shown in Fig. \ref{Fig-NetworkArchitecture}. In this architecture, both the server and client do their own simulations in parallel by running the low-level controller on their devices. Then each player sends its next action to its opponent device for final simulation. Due to floating-point computation errors, the simulations on different devices are very likely to deviate. For solving this issue, server sends world state and score values to the client after its simulation is done and the client will then synchronize itself with the server.

Another important concern in this part was how partial information is handled. Each agent stores opponent's last action spline. Then, when evaluating new candidate splines, opponent's spline is simulated up until a fixed horizon $\tau_{\textit{Opponent Horizon}}$ that is smaller than planning horizon, i.e., $\tau_{\textit{Opponent Horizon}} < \tau_{\textit{Horizon}}$. We did this because it is similar to real life where each character can anticipate movements of other characters by some error.


\section{Evaluations}  \label{Sec-Evaluations}
\subsection{Experimental Setup}
We ran a user study involving $12$ participants to assess the potentials of intelligent middle-level control in our martial arts game prototype. The participants had varying proficiency in video games and martial arts. Screenshots of our $2$-player game are shown in Fig. \ref{Fig-TwoPlayerGameScreenshot} (each player controls the left character on his/her display).

\subsubsection*{Scoring} Each successful punch is rewarded with $1$ to $10$ points depending on its impact. The winner is the first player who gets $100$ points.

\subsubsection*{Slow motion modes} The game was run in slow motion so that players have enough time for planning their movements. We hypothesized that our middle-level control interface could create a strategic \textit{"body chess"} experience instead of a fast-paced action game. However, the camera rotates in real-time speed so the players are able to quickly adjust their point of view to see possible openings for attacks. Since we were not sure what is the right tempo for this game, we tested $3$ different slow motion modes in this study. The chosen speeds for slow motion were $0.12$x, $0.16$x, and $0.2$x. In order to find the best slow motion mode, each pair of players played one match with each slow-motion setting, with ordering of the settings counterbalanced between pairs. For the $6$ pairs, 
each possible ordering was tested once.


\subsubsection*{Goals of User Study} The participants were asked to complete a questionnaire about the most important strengths and weaknesses of the interface. Since there are no games similar to this interface, it was not feasible to conduct a comparative quantitative evaluation. Thus, we designed the questionnaire using open-ended and qualitative questions on the following themes, with the goal of informing future work by both us and other designers and researchers:

\begin{enumerate}
\item What kind of combat techniques does the game allow the players to do?
\item Does the game allow novel gameplay behavior to emerge?
\item How much precision does the low-level controller have in executing players' commands?
\item Which slow motion mode is most suitable for achieving fun gameplay involving high-quality martial arts movements?
\item How can this martial arts interface be improved?
\end{enumerate}

\subsection{Results}
Now we report the results obtained from $12$ participants in our user study. Empty answers and answers that were irrelevant to the asked questions, are not included.

\subsubsection*{Slow Motion Mode} Slow motion modes of $0.12$x, $0.16$x, and $0.2$x were chosen by $1$, $4$, and $7$ people, respectively. The reported reasons are as follows:
\begin{enumerate}
\item The version with $0.12$x speed does not produce fighting experience as the punches do not seem to be effective.
\item In the slowest mode ($0.12$x), it is difficult to anticipate the movements with good precision.
\item In two faster versions ($0.16$x and $0.2$x), it is easier to control the character and react to opponent's moves.
\item The version with $0.2$x speed is chosen by most of the participants because it produces the feeling of action more than other versions.
\end{enumerate}

Our initial belief was that slower versions are easier to work with as they allow more time for planning and anticipation of movements. However, the results suggest that fast-paced gameplay may be more important for martial arts games.

\subsubsection*{Strategies} The main strategies reported by participants are as follows (some participants used more than one strategy):

\begin{enumerate}
\item Staying close and attacking aggressively ($5$ participants).
\item Using one hand for blocking and the other one for punching ($4$ participants).
\item Waiting for opponent to attack and then going for the punch ($2$ participants).
\item Looking for an open side and punching from that side ($1$ participant).
\item Getting hands through the defense of opponent ($1$ participant).
\end{enumerate}

The variety of reported strategies and the gameplay videos suggest that middle-level control provides a good testbed for supporting different styles of gameplay.

\subsubsection*{Movement Precision} $8$ participants reported that character executes the commands with high precision. Other $4$ participants stated that controlling character in slow motion mode is difficult. This suggests that the control algorithm, despite its flaws, is doing a good job in synthesizing dynamic movements, but the control interface was not optimal for all participants.

\subsubsection*{Best Part} The participants were asked to name the best part of their experience while working with the interface. The answers are as follows:
\begin{enumerate}
\item Changing the color of body parts when punching ($3$ participants).
\item "\textit{I like the idea of controlling both hands very much}".
\item "\textit{The way that the hand and body part lit up when punching}".
\item "\textit{The distance between characters can be adjusted in a good way}".
\item "\textit{Seeing the game from the top}".
\item "\textit{Fun to play against a friend}".
\item "\textit{When it comes to punching, the character was quite creative}".
\item "\textit{Different camera angles, realistic approach}".
\item "\textit{Seeing nice landed punches}".
\item "\textit{Nice to win}".
\end{enumerate}

\subsubsection*{Worst Part} The participants have reported the followings as the worst parts of the interface:
\begin{enumerate}
\item Moving arms using mouse drag ($3$ participants).
\item "\textit{Nothing was strikingly bad; but I had some trouble recollecting the right/left-click-equals-right/left-hand rule. At times, I found myself just clicking whatever clicked}".
\item "\textit{It felt like the camera was so close to the body that you almost would like it to be first-person, and especially the camera angle above the head felt like it was from so much above it was not fun to use}".
\item "\textit{Estimating of the time that it takes to hit}".
\item "\textit{When trying to sweep the hands of the opponent away, it wasn't that responsive or intuitive to use}".
\item "\textit{Both parties just end up punching each other, the game is over very quickly, and is not that fun}".
\item "\textit{Unexpected rise of points in opponent's points when in close fight}".
\end{enumerate}

\subsubsection*{Suggestions for Improvement} The participants also made the following suggestions for improving the game:
\begin{enumerate}
\item Adding game-like visual and audio effects ($7$ participants).
\item "\textit{I think it would be cool if you could somehow with mouse make your punches' curves. Maybe dragging the mouse to show the desired curve movement for the hits}".
\item "\textit{Moving the entire body could be possible}".
\item "\textit{Moving hands by clicking and not dragging would make it easier to adjust hand positions}".
\item "\textit{Allowing to crouch which makes it easier to block punches}".
\item "\textit{Allowing to hit arms to incapacitate the other player from blocking punches using them}".
\end{enumerate}


\section{Conclusions} \label{Sec-Conclusions}

We have proposed the concept of intelligent middle-level game control, demonstrated and evaluated through a novel game prototype followed by a user study. This type of game control allows the player to produce novel gameplay through commands that are executed using a low-level controller without using any pre-recorded animations. In our $2$-player martial arts game, each player controls a physically-based simulated character by giving commands such as \textit{"punch in the chest using the left hand"}. Then a low-level controller executes the commands using a real-time control algorithm. Our online continuous control algorithm uses rolling horizon CMA-ES along with custom population seeding techniques.

We evaluated this prototype by running a user study involving $12$ participants. The results show that the interface allows players to come up with various strategies for fighting. The participants have also reported that low-level controller is able to execute the commands with high precision. The users had some difficulties in mastering some of the mechanics such as camera movement and the \textit{Move} task. However, we believe that the interface has potential for further research and novel game experiences.

\section{Limitations and Future Work} \label{Sec-FutureWork}

\subsubsection*{Full-Body Humanoid Characters} Our control algorithm is not currently robust enough to handle full-body humanoid characters in multi-agent environments. We have tested the algorithm successfully in locomotion tasks and single-agent settings. However, heavy perturbations make maintaining balance a very complicated problem. Since using legs is very important in martial arts, enabling this framework to work with full-body humanoid characters is a crucial direction for future work. Full-body characters should also provide enough realism to make the system useful for cognitive, strategic practicing of real martial arts, which we intend to investigate in future user studies.

\subsubsection*{Interface Design} Some reported that they prefer to see the movements in normal speed after they have given a command. However, most of the participants have stated that this character control system is fun and interesting. We are investigating possible ways for improving this interface.

\subsubsection*{Machine Learning} Currently our control algorithm does not apply any kind of machine learning for stabilizing movements. We have already got promising results by adding neural networks on top of our control algorithm in single-agent settings. Our tests show that adding machine learning can be a good approach for reducing sampling budget. Therefore, we believe that using machine learning is one of the low-hanging fruits of this work. On the other hand, being able to function without learning helps as it enables fast iteration when designing and testing the interface design.

\subsubsection*{Other Game Genres} In this work we evaluated intelligent middle-level control only in the context of martial arts games. A reasonable extension to this work is to apply this idea to other games in the sports genre where the quality of movements is important.


\section*{Acknowledgments}

This work has been supported by the Academy of Finland grants $299358$ and $305737$.


\section*{Appendix}
Table \ref{Tbl-ParametersValues} shows the details of all important hyperparameters used in this study.

\begin{table}[h!]
\centering
\caption{Algorithm parameters}
\begin{tabular}{|c|c|} 
\hline
\textbf{Parameter}	& \textbf{Value} \\ \hlineB{3.5}
$\Delta t$  		& $\frac{1}{30}\text{ s}$ \\ \hline
$\tau_{\textit{Max Task}}$  		& $0.5\text{ s}$ \\ \hline
$\tau_{\textit{Horizon}}$  		& $0.6\text{ s}$ \\ \hline
$\tau_{\textit{Opponent Horizon}}$  		& $0.15\text{ s}$ \\ \hline
$n_{\textit{CMA-ES Updates}}$  		& $4$ \\ \hline
$n_{\textit{Population Size}}$  		& $16$ \\ \hline
$n_{\textit{Last Best}}$  				& $3$  \\ \hline
$n_{\textit{Default Pose}}$  			& $3$ \\ \hline
$n_{\textit{Spline Points}}$  		& $3$ \\ \hline
$\sigma_{\textit{Pose}}$  		& $20^\circ$ \\ \hline
$\sigma_{\textit{Move}}$  		& $2\text{ cm}$ \\ \hline
$\sigma_{\textit{Hand Velocity}}$  		& $2.5\text{ m/s}$ \\ \hline
$\sigma_{\textit{Hand Relax Position}}$  		& $1\text{ cm}$ \\ \hline
\end{tabular}
\label{Tbl-ParametersValues}
\end{table}

\bibliographystyle{IEEEtran}
\bibliography{bib_source}

\end{document}